\documentclass[utf8]{frontiersSCNS} 

\setcitestyle{round} 
\usepackage{url,hyperref,lineno,microtype,subcaption}
\usepackage[onehalfspacing]{setspace}

\newcommand{\msun}{M$_\odot$}

\def\keyFont{\fontsize{8}{11}\helveticabold }
\def\firstAuthorLast{Sani {et~al.}} 
\def\Authors{Eleonora Sani\,$^{1,*}$, Federica Ricci\,$^{2,1}$, Fabio La Franca\,$^{2}$, Stefano Bianchi$^{2}$, Angela Bongiorno$^3$, Marcella Brusa$^{4,5}$, Alessandro Marconi$^{6,7}$, Francesca Onori$^8$, Francesco Shankar$^9$, Cristian Vignali$^{4,5}$}
\def\Address{$^{1}$European Southern Observatory, ESO, Santiago de Chile, Chile \\
$^{2}$Dipartimento di Matematica e Fisica, Universit{\`a} degli Studi Roma Tre, Roma, Italy\\
$^3$ INAF Osservatorio Astronomico di Roma, Monteporziocatone, Italy\\
$^4$ Dipartimento di Fisica e Astronomia, Alma Mater Studiorum, Universit{\`a} di Bologna, Bologna, Italy\\
$^5$ INAF Osservatorio Astronomico di Bologna, Italy\\
$^6$ Dipartimento di Fisica e Astronomia, Universit{\`a} degli Studi di Firenze, Firenze, Italy
$^7$ INAF Osservatorio Astrofisico di Arcetri, Firenze, Italy\\
$^8$ Netherlands Institute for Space research, SRON, Utrecht, Netherlands\\
$^9$ Department of Physics and Astronomy, University of Southampton, Highfield, UK
}
\def\corrAuthor{ESO-Chile, Av.da Alonso de Cordova 3107, Vitacura, Santiago de Chile, Chile}

\def\corrEmail{esani@eso.org}

\begin{document}
\onecolumn
\firstpage{1}

\title[Running Title]{NGC~1275: an outlier of the black hole-host scaling relations}

\author[\firstAuthorLast ]{\Authors} 
\address{} 
\correspondance{} 

\extraAuth{}

\maketitle

\begin{abstract}

\section{}
The active galaxy NGC~1275 lies at the center of the Perseus cluster of galaxies, being an archetypal BH-galaxy system that is supposed to fit well with the M$_{BH}$-host scaling relations obtained for quiescent galaxies. Since it harbours an obscured AGN, only recently our group has been able to estimate its black hole mass. \\
Here our aim is to pinpoint NGC~1275 on the less dispersed scaling relations, namely the M$_{BH}$-$\sigma_\star$ and $M_{BH}-L_{bul}$ planes. Starting from our previous work \citep{ricci17b}, we estimate that NGC~1275 falls well outside the intrinsic dispersion of the M$_{BH}$-$\sigma_\star$ plane being $~1.2$~dex (in black hole mass) displaced with respect to the scaling relations. We then perform a 2D morphological decomposition analysis on Spitzer/IRAC images at $3.6~\mu$m and find that, beyond the bright compact nucleus that dominates the central emission, NGC~1275 follows a de Vaucouleurs profile with no sign of significant star formation nor clear merger remnants. Nonetheless, its displacement on the $M_{BH}-L_{3.6,bul}$ plane with respect to the scaling relation is as high as observed in the M$_{BH}$-$\sigma_\star$. \\
We explore various scenarios to interpret such behaviors, of which the most realistic one is the evolutionary pattern followed by NGC~1275 to approach the scaling relation. We indeed speculate that NGC~1275 might be a specimen for those galaxies in which the black holes adjusted to its host. 

\tiny
  \keyFont{ \section{Keywords:} AGN1, AGN2, black hole mass, scaling realtions, infrared, NGC~1275}

\end{abstract}

\section{Overview}
Although active galactic nuclei (AGN) are divided in many flavors, the Unified Model \citep{antonucci93} explains these observational properties with a line-of-sight-dependent scenario, in which a dusty torus makes the emission anisotropic. Nowadays however there is growing observational evidence that type 1 AGN (AGN1) and type 2 (AGN2) are actually characterized by intrinsically different physical properties (e.g., different luminosity functions, \cite{LF05}, \cite{ueda15}; accretion rates, \cite{winter10}; intrinsic X-ray luminosity and black hole (BH) masses, \citep{tueller08}). 
Thanks to our new virial relation based on unbiased physical quantities, i.e. hard X-ray luminosity and Pa$\beta$ emission line FWHM, we have been able to measure, for the first time, with virial methods the supermassive black hole mass (M$_{BH}$) of AGN2 (\cite{LF15}, \cite{LF16}, \cite{onori17a}, whose values have been up today estimated using scaling relations. These relations are calibrated on AGN1 and are unlikely to hold also for all AGN2 \citep{graham08}. With direct virial masses for AGN2, we can  investigate the relation between the M$_{BH}$ and the bulge properties, thus putting a missing piece to the AGN/galaxy coevolution puzzle. 
By selecting unbiased samples of type 1 and type 2 AGN, we have found \citep{onori17a} that in the luminosity range $42.5<\log(L_X/erg~s^{-1})<44.5$, where the two distributions of main AGN types overlap, AGN2 show, on average, significantly smaller Pa$\beta$ and HeI FWHM than AGN1 (1970 km/s instead of 3400 km/s). As expected from the analysis of the FWHM distributions, it results that in this luminosity range, the average M$_{BH}$ of the AGN2 sample ($\log(M_{BH}/M_\odot$) = $7.08\pm 0.10$) is $\sim 0.5$ dex smaller than measured in the AGN1 sample ($\log(M_{BH}/M_\odot$) = $7.61 \pm 0.01$). 
In \cite{ricci17b} we have investigated the M$_{BH}$-$\sigma_\star$ relation and seen how AGN2 harbour less massive BHs than AGN1 at a given velocity dispersion, with BH masses of AGN2 $\sim0.9$ dex smaller than AGN1 at $\sigma_\star\sim 185$~km/s. Equivalently, AGN2 host galaxies have stellar velocity dispersions $\sim0.2$ dex higher than AGN1 hosts at M$_{BH}\sim 10^7$~M$_\odot$. Moreover such feature is not related to the morphological type but could rather be intrinsic. 

To further test this scenario, and to verify whether such a discrepancy holds also for other scaling relations still related to fundamental physical quantities, we are going to investigate the  M$_{BH}$ versus bulge luminosity relation and we present here the case study of NGC~1275. \\
Throughout this paper we assume the standard cosmology with H$_0 = 70$~km/s/Mpc, $\Omega_M= 0.3$, $\Omega_\Lambda=0.7$.

\section{NGC~1275: properties, and BH mass estimate}
NGC~1275 is the brightest and most massive galaxy in the Perseus Cluster (Abell 426), it is a mildly star-forming early-type cD galaxy with a stellar mass of M$_\star=2.43\times10^{11}$~\msun  \citep{mathews06}, i.e the archetypal BH-galaxy system that is supposed to fit well with the M$_{BH}$-host scaling relations obtained for quiescent galaxies. 
By adopting such stellar mass and the M$_{BH}$-M$_\star$ relation of \cite{sani11} (see their Equation 8), one would expect a BH mass of $2.9\times10^8$~M$_\odot$. 
Here we want to verify whether such estimate is consistent with direct measurements and investigate the locus of NGC~1275 over the M$_{BH}$-$\sigma_\star$ and M$_{BH}$-L$_{3.6,bul}$ scaling relations.
Since it also harbour a partially obscured/faint AGN, it represents an excellent laboratory to examine the intrinsic differences between AGN1, AGN2 and quiescent BH. 

A direct M$_{BH}$ measurement by means of molecular gas kinematics is discussed by \cite{wilman05} and more recently by \cite{scharwachter13}, the latter having the advance of a spatial resolution twice higher than the former. According to \cite{scharwachter13} then, the inner (R$\sim 50$~pc) molecular H$_2$ gas kinematics is consistent with a circumnuclear disk and the authors estimate a black hole mass of $8\times10^8$~M$_\odot$, which would be slightly overmassive of a factor of 2 with respect to the M$_{BH}$-M$_\star$ plane taking into account the rms scatter of  the M$_{BH}$-M$_\star$ in \cite{kormendyho13}.
Moreover such estimate is highly uncertain and considered by \cite{scharwachter13} as an upper limit for M$_{BH}$ in NGC~1275. 
Several sources of uncertainties can be identified and are also partly discussed in \cite{scharwachter13}. (I) The spatial resolution should be such to allow a resolution at least 2.5 times the sphere of influence of the BH \citep{hicks10}, while that on IFU data for NGC~1275 can barely relove it. Even though improved by adaptive optics corrections, the resolution of IFU data in \cite{scharwachter13} does not allow to resolve well enought the BH sphere of influence. 
(II) The molecular gas mass in the inner region is non negligible and, just considering it, the BH mass is halved. (III) Since the disk inclination $i$ highly degenerate with M$_{BH}$, it cannot be a free paramer of the kinematical modelling. Unfortunately it can be measured in very few cases: when the disk itself is spatially resolved, e.g. with water maser observations, or when very high S/N IFU data permit to take into account the residual projected velocity on cos $i$ (see Appendix B in \cite{marconi03}).   
This is not the case for \cite{wilman05} and \cite{scharwachter13} and they need to assume the inclination of the circumnuclear disk, whose rotational axis is supposed to be oriented as the axis of the radio jet. This is a strong assumption, as radio jets may not necessarily be aligned perpendicular to the disk. \cite{pastorini07} showed how the broad range of allowed inclinations leads to M$_{BH}$ upper limits rather than realiable measurements. Finally, (IV) the H$_2$ disk kinematics can be significantly affected by bulk motions and non-gravitational forces like gas streams, which are likely to be falling into the core of NGC~1275 \citep{scharwachter13}. \\

As a valid alternative to direct dinamycal measurements, we can use single epoch (SE) scaling relations to estimate M$_{BH}$ in AGN, a method that is widely accepted and not expensive in terms of telescope time \citep{peterson08,shen13}. SE relations are most commonly calibrated by means of either the continuum or broad emission line luminosity and the FWHM of optical emisison line (for a review of these calibrations see \cite{shen13}). 
Recently \cite{koss17} gave estimates based on broad H$\beta$ FWHM and 5100~\AA ~continuum luminosity
, and on broad H$\alpha$ FWHM and integrated luminosity, the latter is considered by the authors as more reliable and leads to M$_{BH}\sim 1.4 \times 10^7$M$_\odot$.  
While such optical SE presciptions are ideal for type 1 AGN, they can be anyhow problematic for a galaxy such as NGC~1275 which is optically classified as Seyfert 1.5  by \cite{ho97}, 
and dimmed optical features produce a high uncertainty in the BH mass estimate. 
For intermediate type 1 Seyfert, the continuum luminosity and the H$_\beta$ complex can be polluted by the FeII emission which is difficult to disentangle, and attenuated by gas extinction; extinction also dims significantly the H$\beta$ emission and partially the H$\alpha$ so that the second estimate is expected to give a higher mass than the first one and a more realiable value. We note that absoption is consistently detected also in X-rays, having NGC~1275 a gas column density of log~N$_H\sim 21.2$ \citep{tueller08}. \\ 
Rather than on the Balmer serie emission lines, we can resort to infrared features to pierce the absorbing gas and have a direct view of the broad line region. Combined with hard X-ray lumonosity, IR features have been used by \cite{ricci17a} to calibrate new and accurate single epoch virial reliations that can be used for obscured or faint AGN.\\ 
With a hard X-ray luminosity of L$_{14-195}=5.13\times10^{43}$~erg/s and a broad Pa$\beta$ with FWHM=2824~km/s \citep{onori17b}, we recently estimated a BH mass of $(2.9\pm0.4)\times10^7$~\msun ~for NGC~1275 \citep{onori17a}.\\ 
Inclination can be a factor of uncertainty also for M$_{BH}$ estimates based on SE scaling relations. The relations depend indeed on the viar factor factor $f$ that account for our ignorance of the morphology, geometry, and kinematics of the BLR. There are evidences for example that more inclined AGN have a larger FWHM of the broad H$_beta$ \citep{bisogni17}, while the f factor decreases with inclination \citep{risaliti11, pancoast14}.  Considered together these results point out a possible disc-like structure of the BLR. Anyhow, the only way to match the single-epoch based M$_{BH}$ with molecular gas kinematic measurements is to increase $f$ of about an order of magnitude, which can happen only under extreme conditions like an Eddington ratio $\lambda_Edd < 0.01$ and inclination $i < 20$~deg (see plot 19 and 20 in \cite{pancoast14}). \\
Our measure is consistent with \cite{koss17} results. Since the two methods are completely independent, we consider SE estimates reliable, while direct kinematical modelling of molecular gas can provide only upper limits of M$_{BH}$ in NGC~1275. The following analysis is threfore based on our value of M$_{BH} \sim 3\times10^7$~\msun. 
Such low value, an order of magnitude lower than what expected for quiescent galaxies and AGN1, is intriguing and deserve further discussion. 

\section{NGC~1275: an outlier of M$_{BH}$-$\sigma_\star$ plane}
\begin{figure}[h!]
\begin{center}
\includegraphics[width=10cm]{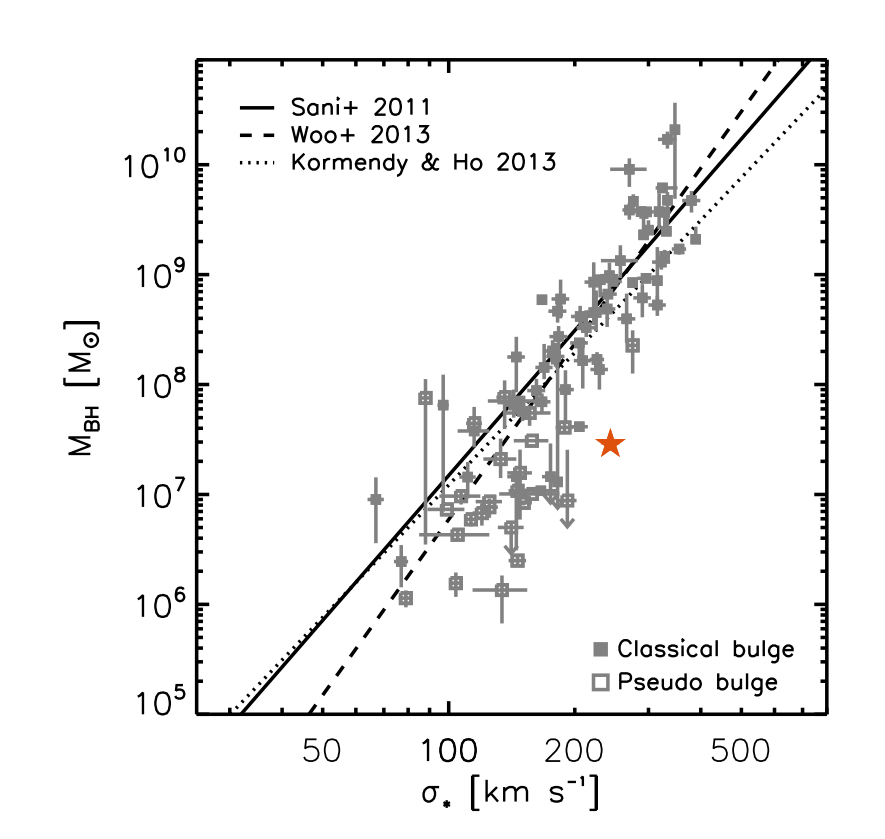}
\end{center}
\caption{The M$_{BH}$-$\sigma_\star$ plane for a local sample of BHs with dynamically measured M$_{BH}$ 
	from \cite{sani11} and \cite{kormendyho13}, scaling relations computed in this two works are also shown. Since our M$_{BH}$ are based on a virial factor $f=4.31$, we also shown the scaling relation by \cite{woo13} that adopts the same $f$ factor. The position of NGC~1275 is marked with an orange star. Classical and pseudo-bulges are marked with filled and open squares respectively.}
\label{fig:1}
\end{figure}

In \cite{ricci17b} we have located type 2 AGN having virial BH mass estimates on the M$_{BH}$-$\sigma_\star$ scaling relation for unobscured and quiescent galaxies. \\
For a median stellar velocity of $\sigma_\star\sim 185$~km/s this analysis shows that AGN2 harbour black holes with a mass $\sim 0.9$~dex smaller than in AGN1 or, equivalently, that host galaxies for type 2 AGN have a hotter kinematics than for type 1, being $\sigma_\star$~0.2~dex higher at M$_{BH}\sim 10^7$~M$_\odot$. Such result does not depend on the host galaxy morphology (i.e. early vs late type).\\
In such framework, NGC~1275 represents the most extreme case as shown in Fig. \ref{fig:1}. For seek of comparison, we show 
only dynamically measured M$_{BH}$. The new sample has been obtained merging the selection of \cite{sani11} with those done in \cite{kormendyho13}, and the velocity dispersion for NGC~1275 has been measured following \cite{ricci17b}. 
 With a stellar velocity dispersion of $\sim240$~km/s, NGC~1275 falls well outside the intrinsic dispersion of the M$_{BH}$-$\sigma_\star$ plane in figure \ref{fig:1} and is indeed $~1.2$~dex displaced with respect to the \cite{sani11, kormendyho13} relations (and 1.35 dex from \cite{woo13}), i.e. $\gtrsim30\%$ more with respect to the average displacement observed for type 2 AGN in \cite{ricci17b}. 
 Several checks against possible biases in the measurements of M$_{BH}$ of AGN2 have been discussed in \cite{onori17a} and no correlation was found between the BLR detectability and infrared flux, nor Xray
flux and luminosity, EW, FWHM, S/N of the spectral features, and not even host orientation or 
column density as measured in the X-rays. The displacement of NGC~1275 with respect to the M$_{BH}$-$\sigma_\star$ plane could be due to an overestimate of the stellar velocity dispersion. 
This is indeed a system composed by the central Perseus galaxy in a late merger state with a smaller spiral galaxy, quite challenging to be interpreted. While indeed an emission-line high velocity component (HVC) is moving towards the dominant system, it lies 57~kpc from the dominant body and cannot affect significantly its bulge kinematics \citep{gillmon04}, on the other hand galaxy merger (independent of the HV system) is still invoked, as an alternative to cooling flow, to explain the present formation of massive and short-lived stars \citep{conselice01}.\\ 
Alternatively, the displacement could be explained by an underestimate of SE M$_{BH}$. We note that to match our single-epoch based M$_{BH}$ with those from 2D gas kinematics so that NGC~1275 lies on the M$_{BH}$-$\sigma_\star$ of qiescent galaxies, the virial $f$ factor should increase of an order of magnitude. Moreover the agreement of our results with \cite{koss17}, hints at an actual undermassive BH in NGC~1275 (see discussion in Sec.~2 for details). \\
To further investigate whether NGC~1275 is an extreme object in terms of BH and host galaxy scaling relations and to verify whether galaxy merging can play a significant role in such puzzle, we dissect mid-infrared images with the goal to measure the bulge luminosity and verify the stellar component morphology. 

\section{NGC~1275: an outlier of the $M_{BH}-L_{3.6,bul}$ plane}
As we aim at performing a detailed photometric bulge decomposition, we choose to analyse mid-infrared images as these wavelengths are tracing both young and old stellar populations. In fact, as demonstrated in \cite{sani11}, the mass-to-light ratio at $3.6~\mu$m doesn't require a color correction to estimate the bulge stellar mass, therefore confirming that the $3.6~\mu$m luminosity is the best tracer of stellar mass yet studied. 
In the following we describe the image analysis of Spitzer data and discuss the location of NGC~1275 with respect to the $M_{BH}-L_{3.6,bul}$ plane. 
\subsection{Spitzer/IRAC data analysis}
We downloaded post-BCD data\footnote{The individual data frames that emerge, calibrated, from the Spitzer pipeline are Level 1, or Basic Calibrated Data, or BCDs. 
The products that come from combining these individual data frames (such as mosaics of individual pointings) are Level 2, or post-BCD, or PBCD data.} from the Spitzer Heritage Archive\footnote{http://irsa.ipac.caltech.edu/applications/Spitzer/SHA/} (SHA) the
deepest and most recent available 3.6~$\mu$m IRAC Astronomical Observation Request (AOR) of 
NGC~1275, 
to allow a reliable 2D decomposition in bulge/disc components 
using the software GALFIT \citep{peng02, peng07}.
Besides the standard inputs [e.g. data, point spread function (PSF) images, etc.], 
GALFIT requires a standard deviation image (which can be directly retrieved from the SHA), used to give relative weights to the pixels during the fit, 
and a bad pixel mask. 
Following the work by \citet{sani11},
we construct a bad pixel frame masking out foreground stars, 
background galaxies and possible irregularly shaped regions such as dust lanes across the galaxy. 
The frames were corrected for geometrical distortion and projected on to a north-east coordinate system with pixel sizes of 1.20 arcsec, equivalent to the original pixels.\\
We fix the background in the fit, estimating it as the mean
surface brightness (with the relative standard deviation) over an annular region surrounding the galaxy between two and three times the optical radius. Foreground sources such as stars or galaxies are not considered in the background calculation by means of a 2.5$\sigma$ rejection criterion. 
We then fitted the image with a bulge+psf model, where the bulge is represented with a Sersic profile and the PSF takes into account the nuclear AGN emission.
As the uncertainties associated to the best-fit parameters are only statistical in GALFIT \citep{haussler07},
and hence underestimate the true uncertainties and degeneracies due to 2D fitting,
we run a grid of 4 fits, varying the Sersic index by $\pm$0.5 with respect to the best fit value, and 
by adding/subtracting to the sky flux its standard deviation. 
In these four grid fits, all the parameters are fixed to the best-fit values obtained in the previous step of fitting, with the exception of 
the magnitude and the effective radius of the Sersic. We then took the 
maximal magnitude variation among the 4 fits as the absolute error. 
With this choice, we overestimate the statistical fit
errors, but carefully constrain the effects due to an uncertain estimate of the background.
The resulting best fit parameters\footnote{To compute the magnitudes at $3.6 \mu$m, we adopt a zero-point of 17.25 in Vega magnitudes, according to the IRAC photometric system \citep{reach05}.} are: 
$m_{3.6,bul}$ = 9.42 $\pm$ 0.21, 
$R_e$ = 42 $\pm$ 15 kpc,
$n_{sers}$ = 4 $\pm$ 0.5 and 
$m_{3.6,psf}$ = 12.12. \\
Figure~\ref{fig:2} shows the two-dimensional analysis of NGC~1275.

\subsection{NGC~1275 and the M$_{BH}$-L$_{3.6,bul}$ plane}

In figure \ref{fig:3} we show the position of NGC~1275 with respect to the M$_{BH}$-L$_{3.6,bul}$ where the data have been taken from \cite{sani11} who also computed the BH-luminosity scaling relation and measured its intrinsic scatter of 0.35~dex. \\
The BH hosted in NGC~1275 is significantly under-massive also with respect to the mass-luminosity plane, as the bulge luminosity measured in the previous section of L$_{3.6,bul}=2.2\times10^{11}$~L$_\odot$ would imply a BH mass $\sim$ 12 times more massive than what measured with single-epoch virial methods. Equivalently, given the BH mass of $2.9\times10^7$~M$_\odot$ the bulge luminosity would be $\sim1.12$~dex lower than what actually measured with our 2D decomposition. \\

\begin{figure}[h!]
\begin{center}
\includegraphics[width=18cm]{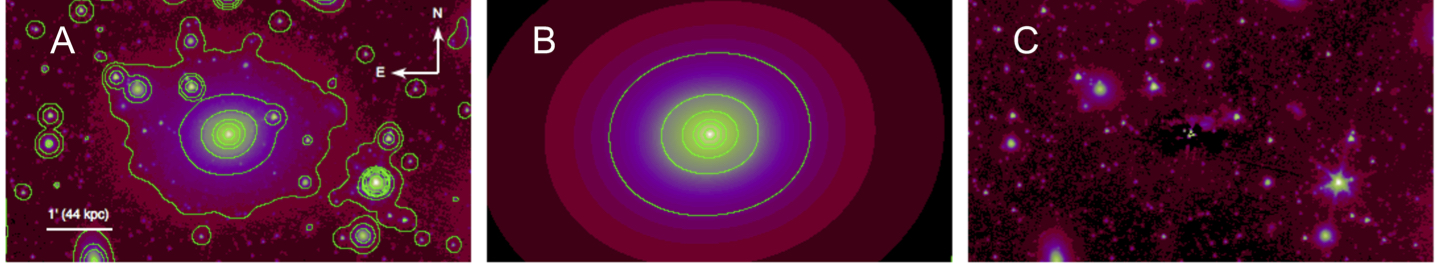}
\end{center}
\caption{Two-dimensional decomposition of Spitzer/IRAC $3.6 mu$m data of NGC1275 performed with GALFIT.The image (panel A), best-fit model (panel B) and residuals (panel C) are shown in logarithmic scale. The residuals are
stretched ($\pm0.25$dex) to highlight the finest details.}
\label{fig:2}
\end{figure}

\begin{figure}[h!]
\begin{center}
\includegraphics[width=10cm]{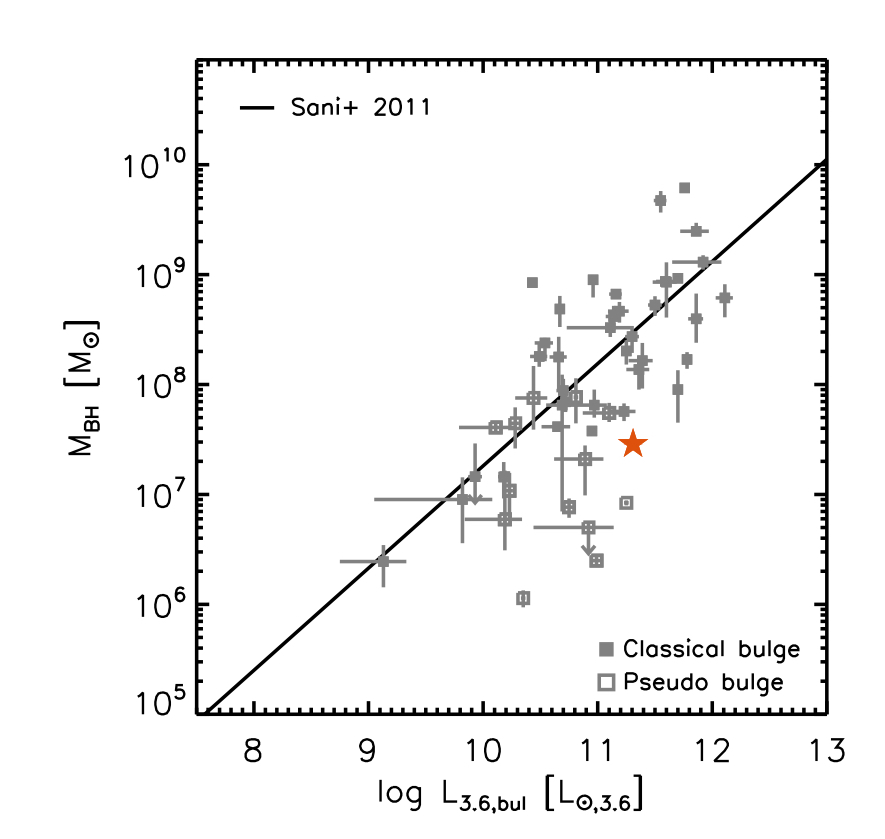}
\end{center}
\caption{The M$_{BH}$-L$_{3.6,bul}$ plane for local sample of  galaxies with a dynamic or maser  measurements of M$_{BH}$, the position of NGC~1275 is marked with an orange star. The scaling relation of \cite{sani11} is also shown for seek of comparison.}
\label{fig:3}
\end{figure}

The galaxy morphology may play an important role to explain this discrepancy. We already mentioned above that NGC~1275 is presumably in a post-merger state, nonetheless the merging process is most probably not altering the bulk of the stellar mass and its distribution. Even though it is classified as an early type galaxy (with morphological type $t=-2$ according to the Hyperleda database, \citep{paturel03}), it is significantly bluer than usual elliptical galaxies. On the other hand, the stellar surface brightness follows a de Vaucouleurs profile at optical wavelengths \citep{mathews06} as well as in the mid-IR. This supports the idea that NGC~1275 is dominated by an old stellar population, as in normal elliptical galaxies, mixed with an additional population of young, luminous stars that does not contribute significantly to the total mass. \\
Our 2D analysis confirms the picture: beyond the bright compact nucleus that dominates the central emission in fig. \ref{fig:2}, star forming regions and dust lanes typical of merging systems are not detected. Nor are detected the mid-IR counterpart of the filaments detected in H$_\alpha$ \citep{conselice01} and X-ray \citep{fabian03} meaning that such filaments are mainly gaseous, and poor in dust content. Additional asymmetric components, as relics of a past merging with a smaller galaxy, might be present and visible in the residuals of the 2D fitting (right panel in fig.~\ref{fig:2}) towards the west/north-west (but the reader should keep in mind the strong stretch of the plot). \\

We note that the displacement from the BH-host planes is consistent with NGC~1275 Seyfert classification. 
Indeed local black holes for which a reliable dynamic measumerement of M$_{BH}$ is possible tend to bias the BH-host scaling relations towards the most massive objects given the requirement to spatially resolve the BH sphere of inluence, and AGN that don't suffer such a bias are actually located below the scaling relations (see figure 3 in \cite{shankar16})
According to our analysis therefore, it is not possible to address the displacement of NGC~1275 from the scaling relations to its morphology. 

The alternative explanation consists in invoking an evolutionary origin. Since their discovery, scaling relations point to a joint galaxy and black hole cosmic evolution. To establish such correlation, one can envisage three possibilities \citep{volonteri12}. ``(\emph{i}) Massive black holes may have grown in symbiosis with their hosts (\emph{ii}), the BH may have dominated the process, with the galaxy catching up later; (\emph{iii}), the galaxy grew first, and the black hole adjusted to its host."\\
It is intriguing to note the position of NGC~1275 in the M$_{BH}$-L$_{3.6,bul}$ plane: it lies below the scaling relation at the edge between massive BHs in classical bulges and in pseudobulges.
When it is possible to populate the M$_{BH}$-L plane with pseudobulges in fact, we can note an interesting behavior: while those with a significant BH mass lie on the scaling relations within the observed scatter,
those with small mass (lower than few $10^7$~M$_\odot$) are significantly displaced \citep{sani11, kormendyho13}.\\ 
We are tempted to interpret the displacement of NGC~1275 from the scaling relation as due to an evolutionary pattern. Even though there is no trace of a pseudobulge from the 2D image analysis, it hosts a relatively small AGN2, and could represent a kind of observational link between secular evolution (thought to be the main mechanism producing pseudobulges) and merger (the builder of massive elliptical galaxies) where the common denominator is the small mass of the black hole, which is obliged by distinct evolutionary processes to adjust to the galaxy. 

\section{Summary and conclusions}
In this work we have investigated the behaviors of NGC~1275, the dominant galaxy of the Perseus cluster, 
in the framework of the M$_{BH}$-host scaling relations. \\
Starting from our previous work \citep{ricci17b}, we have seen how, given the stellar velocity dispersion of $\sim 240$ km/s, NGC~1275 falls well outside the intrinsic dispersion of the M$_{BH}$-$\sigma_\star$ plane being $~1.2$~dex displaced with respect to the scaling relations, i.e. $30\%$ more with respect to the average displacement observed for type 2 AGN in \cite{ricci17b}.\\ 
By means of a 2D morphological decomposition of Spiter/IRAC images at $3.6 \mu$m, we have found that:\\
\begin{itemize}
\item the AGN dust heated emission dominates the nucleus with a magnitude of $m_{3.6,psf}$ = 12.12\\
\item NGC~1275 follows a de Vaucouleurs law extended at large spatial scales with no sign of merger remanence nor of significant star formation activity.\\ 
\item Given a bulge luminosity of L$_{3.6,bul}=2.17\times 10^{11}$~L$_\odot$, NGC~1275 harbours a black holes ~15 times under-massive than what expected for quiescent galaxies that lie within the intrisic scatter of the $M_{BH}-L_{3.6,bul}$ scaling relation.
\end{itemize}
We speculate that the displacement of NGC~1275 with respect to the BH-host scaling relations might 
depend on its evolutionary path where the galaxy grew first, and the black hole is adjusting to its host. We stress that to draw firmer conclusions, it is mandatory to pinpoint on the BH-host planes a statistically significant sample of both type 1 and 2 AGN (Ricci et al. in prep).

\section*{Conflict of Interest Statement}
The authors declare that the research was conducted in the absence of any commercial or financial relationships that could be construed as a potential conflict of interest.

\section*{Author Contributions}
All authors contributed to the interpretation of the observations and the writing of the paper. ES led the analysis, interpretation and wrote the paper. ES, FL, FR defined the strategy. FR performed the 2D modelling. Discussion on the black hole mass measurement has been provided by AM and on scaling relation biases by FS. 

\section*{Funding}
This work has been supported by DGDF ESO funding. 

\section*{Acknowledgments}
This work has been possible thanks to DGDF ESO funding. FR is grateful for the hospitality at by ESO-Chile. ES thanks Dr. Vardha Bennert for fruitfull discussion on the AGN scaling realtions. We acknowledge the usage of the Hyper-Leda database (http://leda.univ-lyon1.fr). This research has made use of the NASA/IPAC extragalactic database (NED).


\bibliographystyle{frontiersinSCNS_ENG_HUMS} 
\bibliography{mybib}
\newpage
\section*{Figure captions}
%
%

\end{document}